\documentclass{emulateapj}

\shorttitle{The narrow-line quasar PMN~J0948+0022}
\shortauthors{The \emph{Fermi}/LAT Collaboration}

\begin{document}


\title{\emph{Fermi}/LAT discovery of gamma-ray emission from a relativistic jet in the narrow-line quasar PMN~J0948+0022}



\author{
A.~A.~Abdo\altaffilmark{1,2}, 
M.~Ackermann\altaffilmark{3}, 
M.~Ajello\altaffilmark{3}, 
M.~Axelsson\altaffilmark{4,5}, 
L.~Baldini\altaffilmark{6}, 
J.~Ballet\altaffilmark{7}, 
G.~Barbiellini\altaffilmark{8,9}, 
D.~Bastieri\altaffilmark{10,11}, 
M.~Battelino\altaffilmark{4,12}, 
B.~M.~Baughman\altaffilmark{13}
% ...
}
% ...
\altaffiltext{1}{National Research Council Research Associate}
\altaffiltext{2}{Space Science Division, Naval Research Laboratory, Washington, DC 20375}
\altaffiltext{3}{W. W. Hansen Experimental Physics Laboratory, Kavli Institute for Particle Astrophysics and Cosmology, Department of Physics and SLAC National Accelerator Laboratory, Stanford University, Stanford, CA 94305}
\altaffiltext{4}{The Oskar Klein Centre for Cosmo Particle Physics, AlbaNova, SE-106 91 Stockholm, Sweden}
\altaffiltext{5}{Department of Astronomy, Stockholm University, SE-106 91 Stockholm, Sweden}
\altaffiltext{6}{Istituto Nazionale di Fisica Nucleare, Sezione di Pisa, I-56127 Pisa, Italy}
\altaffiltext{7}{Laboratoire AIM, CEA-IRFU/CNRS/Universit\'e Paris Diderot, Service d'Astrophysique, CEA Saclay, 91191 Gif sur Yvette, France}
\altaffiltext{8}{Istituto Nazionale di Fisica Nucleare, Sezione di Trieste, I-34127 Trieste, Italy}
\altaffiltext{9}{Dipartimento di Fisica, Universit\`a di Trieste, I-34127 Trieste, Italy}
\altaffiltext{10}{Istituto Nazionale di Fisica Nucleare, Sezione di Padova, I-35131 Padova, Italy}
\altaffiltext{11}{Dipartimento di Fisica ``G. Galilei", Universit\`a di Padova, I-35131 Padova, Italy}
\altaffiltext{12}{Department of Physics, Royal Institute of Technology (KTH), AlbaNova, SE-106 91 Stockholm, Sweden}
\altaffiltext{13}{Department of Physics, Center for Cosmology and Astro-Particle Physics, The Ohio State University, Columbus, OH 43210}
% ...

\begin{abstract}
We report the discovery by the Large Area Telescope (LAT) onboard the \emph{Fermi Gamma-ray Space Telescope} of high-energy $\gamma-$ray emission from the peculiar quasar PMN~J0948+0022 ($z=0.5846$). The optical spectrum of this object exhibits rather narrow H$\beta$ (FWHM(H$\beta$)$\sim 1500$~km~s$^{-1}$), weak forbidden lines and is therefore classified as a narrow-line type I quasar. This class of objects is thought to have relatively small black hole mass and to accrete at high Eddington ratio. The radio loudness and variability of the compact radio core indicates the presence of a relativistic jet. Quasi simultaneous radio-optical-X-ray and $\gamma$-ray observations are presented. Both radio and $\gamma$-ray emission (observed over 5-months) are strongly variable.  The simultaneous optical and X-ray data from \emph{Swift} show a blue continuum attributed to the accretion disk and a hard X-ray spectrum attributed to the jet. The resulting broad band spectral energy distribution (SED) and, in particular, the $\gamma$-ray spectrum measured by \emph{Fermi} are similar to those of more powerful FSRQ. A comparison of the radio and $\gamma$-ray characteristics of PMN~J0948+0022 with the other blazars detected by LAT shows that this source has a relatively low radio and $\gamma$-ray power, with respect to other FSRQ. The physical parameters obtained from modelling the SED also fall at the low power end of the FSRQ parameter region discussed in Celotti \& Ghisellini (2008). We suggest that the similarity of the SED of PMN~J0948+0022 to that of more massive and more powerful quasars can be understood in a scenario in which the SED properties depend on the Eddington ratio rather than on the absolute power. 
\end{abstract}


\section{Introduction}
It is now widely recognised that strong radio sources associated with active galactic nuclei (AGN) must be powered by collimated relativistic energy flows (Rees 1966). The bulk Lorentz factors ($\Gamma$) of these flows may be different in different systems and at different distances from the active nucleus. If a blob of plasma moving at relativistic speed is observed at small angles to the jet axis ($\theta \leq 1/\Gamma$) the observed radiation is amplified and the timescales shortened due to relativistic effects. Such systems are generically called blazars (Blandford \& Rees 1978). Blazars have been often classified in subcategories: Flat-Spectrum Radio Quasars (FSRQ), characterized by strong and broad optical emission lines, and BL Lac objects, when no emission lines are apparent above the optical/UV continuum (e.g. Urry \& Padovani 1995).

Remarkably, the \emph{Compton Gamma-Ray Observatory} (CGRO) together with the first generation of Cherenkov Telescopes discovered that the SED of a number of the brightest blazars extend to the $\gamma$-ray range, showing two broad components: the first one, covering radio to soft X-rays, is thought to be due to the synchrotron emission from relativistic electrons, while the second one, covering the hard X-/$\gamma-$ray band, is generally attributed to inverse-Compton (IC) emission. The seed photons for the IC process can originate from the synchrotron radiation itself (synchrotron self-Compton, SSC; e.g. Ghisellini et al. 1985) or from an external source, like the accretion disk, the broad-line region or a dusty torus (external Compton, EC; e.g. Dermer et al. 1992, Sikora et al. 1994, 
B{\l}a\.zejowski et al. 2000).
 
Compiling and averaging the SED of the brightest blazars, Fossati et al. (1998) found an interesting trend -- the 
blazar sequence -- whereby for sources with low bolometric luminosity both components peak at high frequencies (UV/soft X-rays for synchrotron and TeV for IC -- High-frequency peaked BL Lacs, HBL), while, for increasing luminosities, both 
peaks shift to lower frequencies (Low-frequency peaked BL Lacs, LBL and FSRQ).

Ghisellini et al. (1998) proposed to explain the sequence in terms of correlation between the random Lorentz factor of electrons emitting at the peaks of the SED ($\gamma_{\rm peak}$) and the global energy density ($U$) in the comoving frame. HBL have low $U$ and high $\gamma_{\rm peak}$, while FSRQ have high $U$ and low $\gamma_{\rm peak}$. The sequence can also be interpreted in an evolutionary frame (B\"ottcher \& Dermer 2002, Cavaliere \& D'Elia 2002).

It is important to stress that the selection of objects with which the sequence was constructed was admittedly biased by the available samples, within which only a limited number of objects had $\gamma$-ray data (Maraschi \& Tavecchio 2001). In fact challenges have been raised to the validity of the sequence (for a review, see Padovani 2007 and references therein), which however could be overcome (Maraschi et al. 2008, Ghisellini \& Tavecchio 2008). Today, the ongoing \emph{Fermi} mission is expected to provide a deeper and unbiased survey of the whole $\gamma-$ray sky, compared to that available during the \emph{CGRO}/EGRET era, yielding possible surprises, as we will show in the present work.

Although the origin of relativistic jets is presently still not understood, there is increasing evidence that the properties of jets are related to the properties of the accretion flow which feeds the central black hole. In the case of stellar mass black holes the observed phenomenology points to an association of jet launching  with accretion ``modes'' characterized by different spectral and timing properties (Fender \& Belloni 2004). In the extragalactic domain the separation of radio sources into two broad classes (FRI and FRII), as well as the properties of their respective ``beamed'' representatives, BL Lacs and FSRQ, can be basically understood within a scenario based on the accretion mode: in the first class accretion onto the central black hole is sub-critical (in Eddington units) leading to radiatively inefficient accretion flows and relatively weak jets, while in the second one the accretion rate is near critical, giving rise to bright disks and powerful jets (Ghisellini \& Celotti 2001, Maraschi 2001, Maraschi \& Tavecchio 2003; see, however, also Blandford \& Levinson 1995).

In this respect, the case of radio-loud narrow line Seyfert 1 (NLS1) active nuclei has received increasing attention. NLS1 are characterized by an optical spectrum with narrow permitted lines FWHM(H$\beta$)~$< 2000$ ~km/s, the ratio between [OIII]$\lambda 5007$ and H$\beta$ smaller than 3 and a bump due to FeII (see, e.g., Pogge 2000 for a review). They exhibit also prominent soft X-ray excesses. These properties point to very high (near Eddington) accretion rates and relatively low masses ($10^6 - 10^8 M_{\odot}$) (Boroson 2002; see, however, Decarli et al. 2008, Marconi et al. 2008). Only a small percentage of NLS1 are radio-loud ($RL = (S_{\nu=4.85\rm GHz}/S_{\nu=440\rm nm}) > 10$) or very radio-loud ($RL>100$) (7\% and 2.5 \% respectively). Their flat radio spectra suggest that several of them could host relativistic jets: in fact VLBI variability indicates extremly high brightness temperatures and in some cases superluminal expansion has been observed (Komossa et al. 2006, Doi et al. 2006). 
 
Recently, Yuan et al. (2008) studied a complete sample of radio-loud NLS1 selected from the \emph{Sloan Digital Sky Survey} (SDSS) sample. They find that a large fraction of those for which X-ray data exist show broad band spectra similar to those of HBL, with peaks close to the UV band. The study of simultaneous optical/UV/X-ray data of a sample of radio-loud NLS1 revealed that these sources often display a hard X-ray component, especially in bright optical and ultraviolet states, thus supporting the possible contribution of a relativistic jet, in some cases similar to FSRQ (Foschini et al. 2009). 

This class of sources thus is of extreme interest for extending the studies of the properties of relativistic jets to different mass and power scales. While it is clear that the most radio-loud NLS1 should host relativistic jets, the properties of such jets at high energy are essentially unknown. Observations of a few selected sources in the TeV energy range with the \emph{Whipple} and \emph{HESS} Cerenkov telescopes were performed, but yielded only upper limits (Falcone et al. 2004, Aharonian et al. 2008). A detection at high energy (GeV or TeV $\gamma$-rays) is essential to complete the knowledge about the SED, allowing to constrain the inverse-Compton parameters and discuss analogies and differences with previously known blazars.
 
Here we present the first detection, by the \emph{Fermi}/LAT, of $\gamma-$rays from one of these radio-loud NLS1, the quasar PMN~J0948+0022 ($z=0.5846$). This makes it possible to build the first whole SED from radio to $\gamma-$rays of a radio-loud NLS1, to constrain the inverse-Compton emission, and to evaluate the role of this new type of source in the framework of the blazar sequence and evolution. The paper is organized as follows: after a short identity card of the source in Sect. 2, the analysis of \emph{Fermi}/LAT, \emph{Swift}, Effelsberg and Owens Valley Radio Observatory data is presented in Sect.~3; Sect.~4 deals with the SED and the model selected to fit to the broad-band spectrum, while the discussion and conclusions are given in Sect.~5. Throughout this work, we adopted a $\Lambda$CDM cosmology from the most recent \emph{WMAP} results, which give the following values for the cosmological parameters: $h = 0.71$, $\Omega_m = 0.27$, $\Omega_\Lambda = 0.73$ and with the Hubble-Lema\^{i}tre constant $H_0=100h$ km s$^{-1}$ Mpc$^{-1}$ (Komatsu et al. 2009).

\section{The source PMN~J0948+0022}
First identified in the MIT-Green Bank radio survey at 5~GHz (Bennett et al. 1986), this quasar is part of the sample drawn from the SDSS and FIRST (\emph{Faint Images of the Radio Sky at Twenty-Centimeters}) in a systematic search for radio loud NLS1 objects (Zhou et al. 2003, Komossa et al. 2006, Zhou et al. 2006, Yuan et al. 2008). The flux ratio [OIII]/H$\beta = 0.1 < 3$ (according to the most recent measurements by Zhou et al. 2006) indicates that the Balmer lines are indeed originated from the usual broad-line region, meaning that this is not a Type II AGN (Zhou et al. 2003). There is no obscured broad-line region, as supported by the absence of any additional absorption in the optical (see the analyses by Zhou et al. 2003, 2006) and X-ray spectra (no additional $N_{\rm H}$ is required in the fit of \emph{Swift}/XRT data in the present work; see Sect. 3.2). A reanalysis of SDSS data by Zhou et al. (2006) succeeded in separating the H$\beta$ into a broad and narrow components (see also Rodr\'iguez-Ardila et al. 2000), where the ``broad'' component has FWHM~$= 1432 \pm 87$~km~s$^{-1}$. Therefore, PMN~J0948+0022 has a broad and a narrow line region, but the FWHM of the ``broad'' lines is small, less than $2000$~km~s$^{-1}$. 

Its strong and variable radio emission ($RL > 1000$), together with a flat and inverted radio spectrum ($\alpha_{r}=-0.24$, with $\alpha$ defined as $S_{\nu}\propto \nu^{-\alpha}$), make PMN~J$0948+0022$  one of the few undoubtely radio-loud narrow-line Seyfert 1 quasars. In addition, VLBI observations at different epochs revealed high brightness temperatures and significant flux density variations, requiring a Doppler factor $\delta > 2.5$ and a viewing angle $<22^{\circ}$ (Doi et al. 2006). Now, a $\gamma-$ray detection is reported in the first list of bright $\gamma$-ray sources (significance $>10\sigma$) detected by \emph{Fermi}/LAT (Abdo et al. 2009a) and, specifically, in the List of Bright AGN Sources (LBAS, Abdo et al. 2009b).

\begin{figure}
\centering
\includegraphics[scale=0.5]{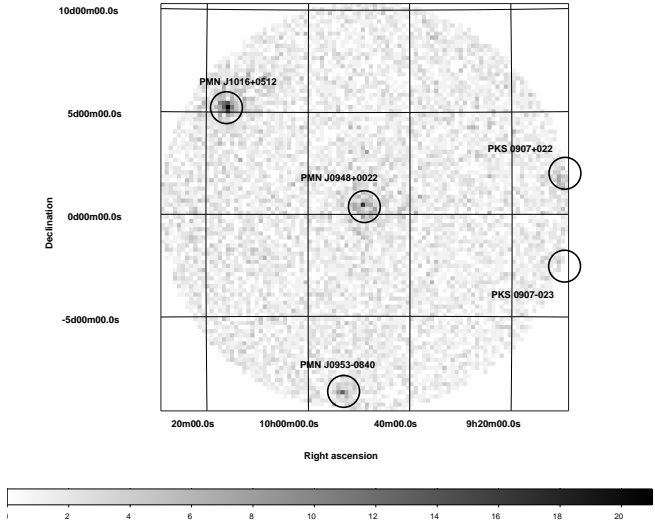}
\caption{\emph{Fermi}/LAT counts map ($E> 200$~MeV) of the region centered on PKS~J0948+0022 with radius $10^{\circ}$. The pixel size is $0^{\circ}.2$. The gray scale bar is in units of LAT counts integrated in the 5-month period. Epoch of coordinates is J2000. Nearby sources included in the likelihood analysis are also indicated.}
\label{fig:latmap}
\end{figure}

\begin{figure*}
\centering
\includegraphics[angle=270,scale=0.5]{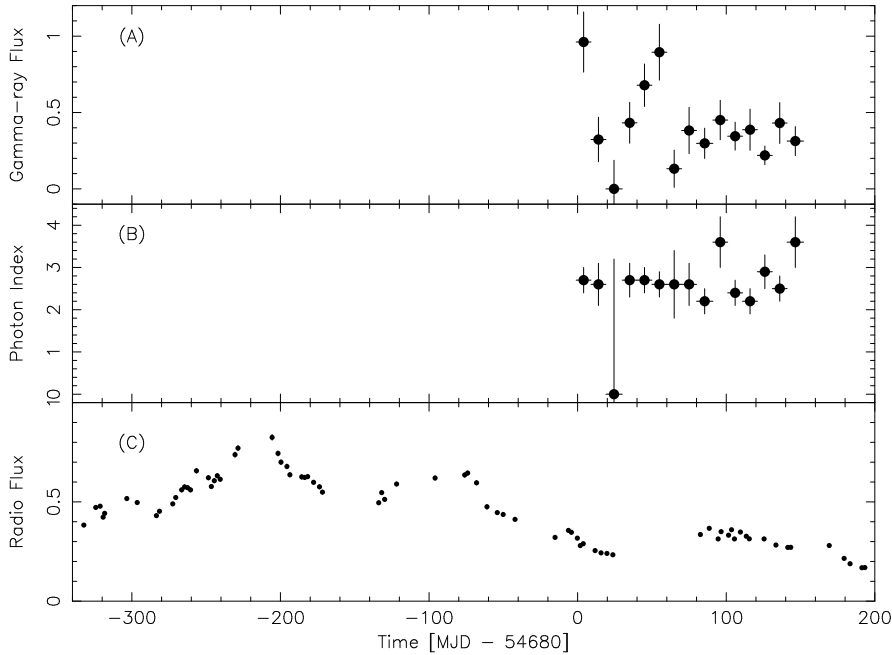}
\caption{(\emph{panel A}) \emph{Fermi}/LAT flux for $E>200$~MeV in units of [$10^{-7}$~ph~cm$^{-2}$~s$^{-1}$] with $10$-day time bins and (\emph{panel B}) the corresponding photon index. (\emph{panel C}) OVRO radio flux at 15 GHz [Jy]. Time scale is from 4 September 2007 at 16:25:41 UTC to 11 February 2009 06:53:48 UTC and the day 0 in the X-axis refers to the beginning of LAT observations (1 August 2008).}
\label{fig:lcurves}
\end{figure*}

\section{Data Analysis}

\subsection{Fermi/LAT}
PMN~J0948+0022 appears in the LBAS derived from the first three months of the \emph{Fermi}/LAT all-sky survey. It is associated (93\% confidence level) with the source 0FGL~J0948.3+0019 (Abdo et al. 2009a,b). In order to improve the positional accuracy for this source and hence increase the confidence level of the association, we added two more months of data to the three months of the LBAS, thus providing 5 months of data (from August through December 2008).

The data from the Large Area Telescope (LAT, Atwood et al. 2008) were analyzed using the \emph{Fermi}/LAT {\tt Science Tools v 9.8.2}, a specific LAT software package\footnote{http://fermi.gsfc.nasa.gov/ssc/data/analysis/software/}. In summary, we performed the analysis using the guidelines described in detail in the LBAS paper (Abdo et al. 2009a,b), but with some specific features to best fit the characteristics of this source. Events of ``Diffuse'' class in the \texttt{Science Tools}, coming from zenith angles $< 105^\circ$ (to avoid Earth's albedo) were extracted from a region with a $10^{\circ}$ radius centered on the coordinates of the radio position of PMN~J0948+0022 ($RA=09^{\rm h}48^{\rm m}57^{\rm s}.6$ and $Dec=+00^{\circ}22'12''.0$, J$2000$). Because of calibration uncertainties at low energies, data were selected with energies above $200$~MeV, with no $\gamma-$ray detected at energies $> 3$~GeV. The $\gamma-$ray background is mainly due to three components: the diffuse emission from the Milky Way, the diffuse extragalactic background and  the instrumental background. All these components were modeled and accounted for in the analysis.

An unbinned likelihood algorithm, implemented in the \texttt{LAT Science Tools} as the \texttt{gtlike} task, was used to analyze the data. PMN~J0948+0022 was modelled together with four other nearby sources: PKS~0907+022, PKS~0907-023, PMN~J1016+0512, PMN~J0953-0840, to take possible contaminations into account, due to the strongly energy-dependent point-spread function (see Figure~\ref{fig:latmap}).
 
The best fit position of the LAT source, after 5 months of integration, is $RA=09^{\rm h}48^{\rm m}58^{\rm s}$ and $Dec=+00^{\circ}22'48''$ (J$2000$), with a 95\% error radius of $12'36''$ ($0^{\circ}.21$) and detection test statistics $TS = 305$ (a $17\sigma$ detection; see Mattox et al. 1996 for the definition of $TS$). Figure~\ref{fig:latmap} shows the LAT counts map with the peak of counts consistent with the radio position of PMN~J0948+0022. The Figure-of-Merit (FoM) method to associate $\gamma-$ray sources with known radio counterparts was used (Sowards-Emmerd et al. 2003; Abdo et al. 2009a,b). The value of the FoM increased from 18.64 and 93\% confidence level (Bright AGN List, 3 months data) to 53.92 and 99\% probability that the association is correct (5 months data).

Fig.~\ref{fig:lcurves}~(\emph{A}) displays the light curve for $E>200$ MeV integrated over $10$-day time bins between $1$~August~$2008$, $00$~UT, to $1$~January~$2009$, $00$~UT (MJD 54679 - 54832). The corresponding photon indices $\Gamma$, defined as $F(E) \propto E^{-\Gamma}$, are shown in Fig.~\ref{fig:lcurves}~(\emph{B}). The light curve indicates clear variability of a factor of $\sim 10$ over a timescale of weeks, confirmed with a $\chi^2=34.5$ for $14$ degrees of freedom ($\tilde{\chi}^2=2.5$) for a fit with a constant, and an excess variance of $0.19\pm 0.06$ (see Nandra et al. 1997 for a definition of excess variance). On the other hand, the photon index does not show evident variability and can be fitted with a constant ($\tilde{\chi}^2=0.80$). The radio data at $15$~GHz in Fig.~\ref{fig:lcurves}~(\emph{C}), described in Section~3.3, displays no variability during the same time interval. We note, however, that radio data are missing during the episode of variability at $\gamma$-rays at the beginning of the LAT lightcurve (Fig.~\ref{fig:lcurves}~(\emph{A})) and a strong radio outburst was observed well before the launch of \emph{Fermi}.

The spectrum averaged over the whole data set was initially fitted with a single power-law model with $\Gamma = 2.6\pm 0.1$ and flux ($E>200$~MeV) equal to $(4.0\pm 0.3)\times 10^{-8}$~ph~cm$^{-2}$~s$^{-1}$ ($TS = 298$). A fit with a broken-power law model gives a slight increase of the $TS$. The spectral parameters are the following: the break energy is $E_{\rm b} = 1.0 \pm 0.4$~GeV, while the photon index for $E<E_{\rm b}$ is $\Gamma_1=2.3\pm 0.2$ and for $E>E_{\rm b}$ is $\Gamma_2=3.4\pm 0.5$. The integrated flux ($E>200$~MeV) is $(3.9\pm 0.3)\times 10^{-8}$~ph~cm$^{-2}$~s$^{-1}$ ($TS = 305$). The likelihood test ratio gives 97\% probability in favor of the broken power-law model with respect to the single power-law. 

It is worth noting that the quoted errors are statistical only. Systematic errors should be added. According to the studies on the Vela Pulsar (Abdo et al. 2009c), our current conservative estimates of systematic errors are $<30$\% for flux measurements and $0.1$ for the photon index. Significant reduction of such systematic uncertainties is expected once the calibration of the LAT instrument is completed. 

\begin{table*}
\begin{center}
\scriptsize
\caption{Summary of results from analysis of the \emph{Swift} data obtained on $5$ December $2008$ (ObsID $00031306001$). See the text for details.\label{tab:swift}}
\begin{tabular}{cccccc}
\tableline
\tableline
\multicolumn{6}{c}{BAT ($20-100$~keV)}\\
\tableline
\multicolumn{3}{c}{Exposure} & \multicolumn{3}{c}{Flux$_{20-100 \rm keV}$}\\
\multicolumn{3}{c}{[ks]}     & \multicolumn{3}{c}{[$10^{-10}$~erg~cm$^{-2}$~s$^{-1}$]}\\
\tableline
\multicolumn{3}{c}{$4.3$}  & \multicolumn{3}{c}{$< 7.0 $} \\ 
\tableline
\multicolumn{6}{c}{XRT ($0.2-10$~keV)}\\
\tableline
Exposure & $N_{\rm H}$(\tablenotemark{*}) & $\Gamma$ & Normalization at 1 keV & Flux$_{2-10 \rm keV}$ & $\chi^{2}$/dof \\
{[ks]} & [$10^{20}$~cm$^{-2}$] & {} & [$10^{-3}$~ph~cm$^{-2}$~s$^{-1}$~keV$^{-1}$] & [$10^{-12}$~erg~cm$^{-2}$~s$^{-1}$] & {} \\
\tableline
$4.2$ & $5.22$ & $1.83\pm 0.17$ & $1.5\pm 0.3$ & $2.2\pm 0.2$	& $5.1/8$\\
\tableline
\multicolumn{6}{c}{UVOT (observed magnitudes)}\\
\tableline
$v$ & $b$ & $u$ & $uvw1$ & $uvm2$ & $uvw2$ \\
{[$5468$~\AA]} & {[$4392$~\AA]} & {[$3465$~\AA]} & {[$2600$~\AA]} & {[$2246$~\AA]} & {[$1928$~\AA]} \\
\tableline
$18.2\pm 0.1$ & $18.56\pm 0.07$ & $17.79\pm 0.06$ & $17.48\pm 0.06$ & $17.50\pm 0.06$ & $17.55\pm 0.05$\\
\tableline
\end{tabular}
\tablenotetext{*}{Fixed value from measurements of the Galactic absorption by the Leiden-Argentine-Bonn survey (Kalberla et al. 2005).}
\normalsize
\end{center}
\end{table*}

\subsection{Swift}
On $5$~December~$2008$ at 02:25~UTC (MJD 54805.10), \emph{Swift} observed PMN~J$0948+0022$ (ObsID~$00031306001$, exposure $\sim 4$~ks). For the screening, reduction and analysis of the data from the three instruments (BAT, Barthelmy et al. 2005; XRT, Burrows et al. 2005; UVOT, Roming et al. 2005) onboard the \emph{Swift} satellite (Gehrels et al. 2004), we used the \texttt{HEASoft v. 6.6.1} software package, together with the \texttt{CALDB} updated on $9$~December~$2008$. 

The X-Ray Telescope XRT (0.2-10~keV energy band) was used in photon counting mode and no evidence of pile-up was found. Data were processed and screened by using the \texttt{xrtpipeline} task with default parameters and grades 0-12 (the single to quadruple pixels events). The spectrum was rebinned to have at least $30$ counts per bin. 

No detection was found at $E>10$~keV with BAT, after having binned, cleaned from hot pixels, deconvolved and integrated all the data available in this pointing. 

UVOT observed the source with all the six available filters. Data were integrated with the \texttt{uvotimsum} task and then analyzed by using the \texttt{uvotsource} task, with a source region radius of $5''$ for the optical filters and $10''$ for the ultraviolet, while the background was extracted from a circular region $1'$-sized and centered in a nearby source-free region. It was not possible to select an annular region centered on PMN J0948+0022, because of nearby sources. The observed magnitudes were dereddened according to the extinction laws of Cardelli et al. (1989) with $A_{V}=0.277$ and then converted into flux densities according to the standard formulae and zeropoints (Poole et al. 2008). 

The results are summarized in Table~\ref{tab:swift}.

\begin{figure}
\centering
\includegraphics[angle=270,clip,trim=100 0 100 0,scale=0.4]{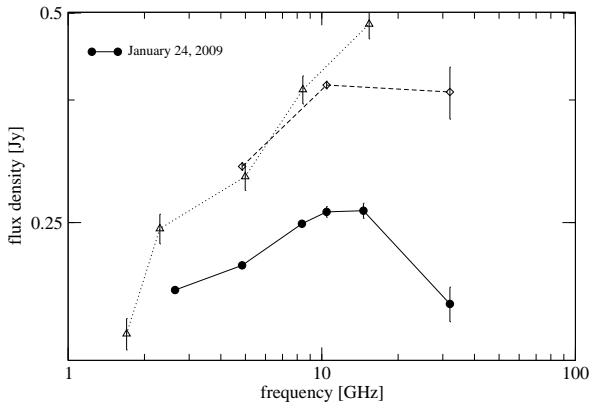}
\caption{The evolution of the radio spectrum of PMN~J0948+022. Filled circles denote the Effelsberg observations of January 2009. Archival data is shown in grey: triangles represent VLBA measurements conducted in October 2003 (Doi et al.2006), diamonds represent archival Effelsberg measurements obtained in 2006 (Vollmer et al. 2008).}
\label{radio}
\end{figure}

\subsection{Radio}
\subsubsection{Effelsberg}

The centimeter spectrum of PMN J0948+0022 was observed with the Effelsberg 100~m telescope on 24 January 2009 (MJD 54855.5) within the framework of a \emph{Fermi} related monitoring program of potential $\gamma$-ray blazars (F-GAMMA project, Fuhrmann et al. 2007). The measurements were conducted with the secondary focus heterodyne receivers at 2.64, 4.85, 8.35, 10.45, 14.60 and 32.00~GHz. The observations were performed quasi-simultaneously with cross-scans, that is slewing over the source position, in azimuth and elevation direction, with adaptive numbers of sub-scans for reaching the desired sensitivity (for details, see Fuhrmann et al. 2008; Angelakis et al. 2008). Pointing off-set correction, gain correction, atmospheric opacity correction and sensitivity correction have been applied to the data.

The acquired radio spectrum has a convex shape with a turnover frequency between 10.45 and 14.60~GHz (Fig.~\ref{radio}). The low-frequency part spectral index $\alpha_{2.64}^{10.45}$, measured between 2.64 and 10.45~GHz, is $-0.18\pm0.02$, whereas the high-frequency optically thin spectral index $\alpha_{14.6}^{32}=0.39$.

The comparision of the acquired spectrum with previously observed ones (Vollmer et al. 2008, Doi et al. 2006) reveals that the source is presently in a much lower flux density state (see Fig. \ref{radio}). This indicates intense variability. From the change of the 4.85~GHz flux density over about three years (Vollmer et al. 2008) we estimate a variability brightness temperature (e.g Fuhrmann et al. 2008) of $1.4\times10^{11}$~K. Assuming the equipartition brightness temperature limit of $\sim10^{11}$~K (Readhead et al. 1994), we obtain a lower limit for the Doppler factor of $\delta>1.4$.

\subsubsection{Owens Valley Radio Observatory (OVRO)}
PMN~J0948+0022 has been observed regularly from 4 September 2007 at 16:25 UTC to 11 February 2009 06:53 UTC (MJD 54347.68-54873.29) at 15~GHz by the Owens Valley Radio Observatory (OVRO) 40~m telescope as part of an ongoing \emph{Fermi} blazar monitoring program of all 1159 CGRaBS blazars north of declination $-20$ degrees (Healey et al. 2008).  Flux densities were measured using azimuth double switching as described in Readhead et al. (1989).  The relative uncertainties in flux density result from a 5~mJy typical thermal uncertainty in quadrature with a 1.6\% systematic uncertainty.  The absolute flux density scale is calibrated to about 5\% using the model for 3C~286 by Baars et al. (1977).  This absolute uncertainty is not included in the plotted errors.

PMN~J0948+0022 has been reported to show variability by a factor of 2 in radio over year time-scales (Zhou et al., 2003) and $\sim$~31\% fluctuations in month time-scales (Doi et al., 2006).  The OVRO 40~m 15~GHz time series shows clear structure at timescales down to weeks, and year-scale fluctuations by a factor of 4 (Fig. \ref{fig:lcurves}, \emph{panel C}). The rapid variability we observe in this object -- at least 400~mJy in 77~days or 5~mJy/day -- enables us to determine a variability brightness temperature of $\sim 2\times 10^{13}$~K, assuming the $\Lambda$CDM cosmology described in Sect.~1.  

It is often not easy to determine the optically thin spectral index of blazars at radio frequencies because they are complex structures, with different regions becoming optically thin at different radio frequencies. In the present case the most recent results show a turnover between 10 and 15~GHz, and a $15-30$~GHz spectral index of $\sim 0.4$, but we do not believe that this is the optically thin spectral index. It is much more likely that one is still seeing synchrotron self-absorption so that the spectrum between 15 GHz and 30 GHz is much flatter than the true optically thin spectral index. In such cases, it is safer to assume an optically thin spectral index of $0.75$ and to use the frequency of observation. These only have a small effect on the derived $T_{\rm eq}$ unless $\alpha$ is very close to $0.5$, which is too flat, in our view, for an optically thin spectral index in most cases.

The equipartition brightness temperature (Readhead, 1994), in the current cosmological model, is then $T_{\rm eq} \sim 5.5\times 10^{10}$~K, assuming an average optically thin spectral index of $0.75$, and hence the equipartition Doppler factor is $\delta \sim 7$, which is typical of highly variable blazars.  This agrees with Doi et al. (2006), who reported an equipartition Doppler factor $\delta > 2.7$, and the Effelsberg lower limit reported in Sect.~3.3.1.  This suggests that, in the compact radio emission regions in this object, the Lorentz $\Gamma$ factor is of order 10.  

\begin{figure*}
\centering
\includegraphics[scale=0.8,clip,trim=0 40 0 0]{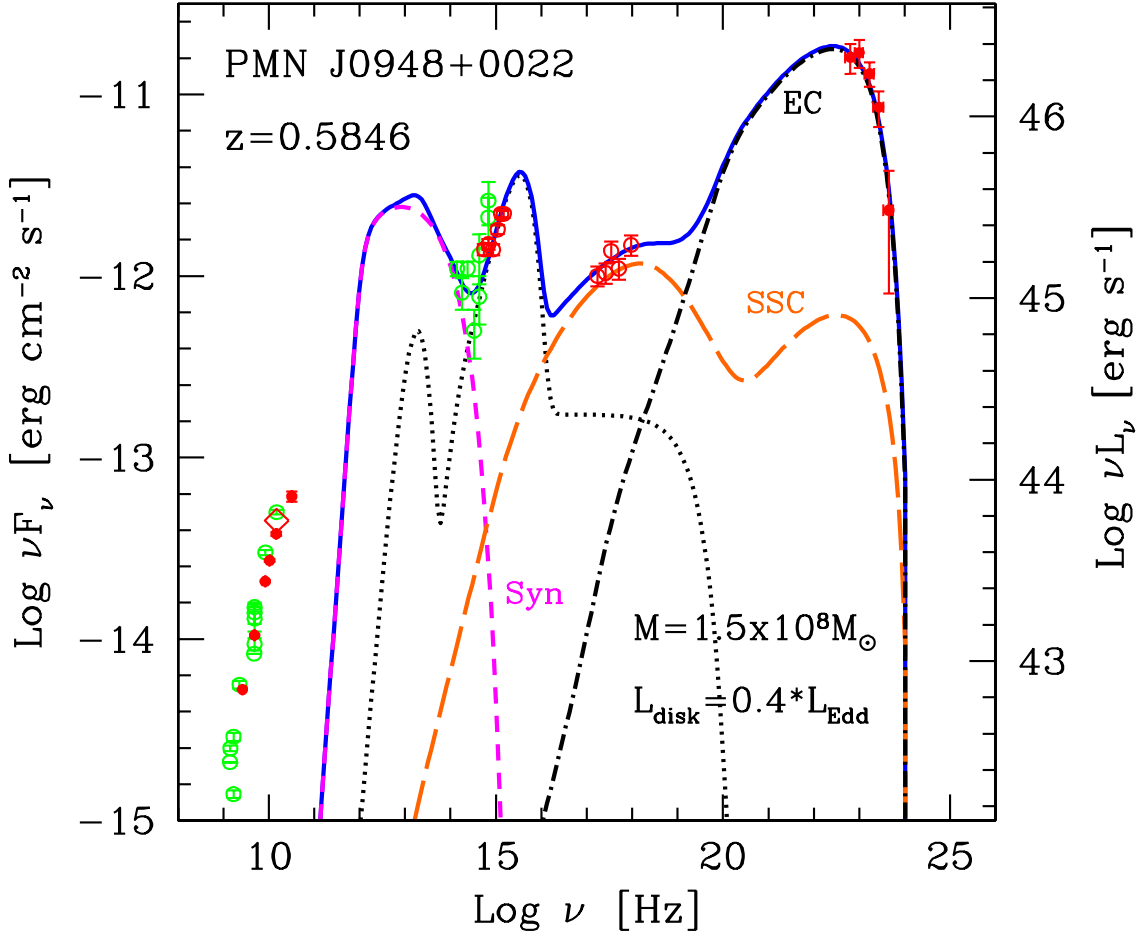}
\caption{Spectral Energy Distribution of PMN~J0948+0022. \emph{Fermi}/LAT (5-months data); \emph{Swift} XRT and UVOT (5 December 2008); Effelsberg (24 January 2009) and OVRO (average in the 5-months of LAT data, indicated with a red diamond) are indicated with red symbols. Archival data are marked with green symbols. Radio data: from $1.4$ to $15$~GHz from Bennett et al. (1986), Becker et al. (1991), Gregory \& Condon (1991), White \& Becker (1992), Griffith et al. (1995), Doi et al. (2006). Optical/IR: USNO B1, $B$, $R$, $I$ filters (Monet et al. 2003); 2MASS $J$, $H$, $K$ filters (Cutri et al. 2003). The dotted line indicates the contributions from the infrared torus, the accretion disk and the X-ray corona. The synchrotron (self-absorbed) is shown with a small dash line. The SSC and EC components are displayed with dashed and dot-dashed lines, respectively. The continuous line indicates the sum of all the contributions.}
\label{sed}
\end{figure*}

\section{Spectral Energy Distribution (SED)}

Fig.~\ref{sed} displays the spectral energy distribution (SED) built with the \emph{Fermi}/LAT, 
\emph{Swift}, and Effelsberg and OVRO data analyzed in the present work (red symbols) together 
with archival data (green symbols). 
Archival radio data are from Bennett et al. (1986), Becker et al. 
(1991), Gregory \& Condon (1991), White \& Becker (1992), Griffith et al. (1995), Doi et al. (2006); 
optical/IR data are from USNO B1 for $B$, $R$, $I$ filters (Monet et al. 2003) and from 2MASS for 
$J$, $H$, $K$ filters (Cutri et al. 2003).

The resulting SED strongly resembles that of a typical high power blazar, with two non--thermal 
emission peaks in the far IR and between $10^{22}-10^{23}$~Hz ($40-400$~MeV). 
Also the peak produced by the accretion disk is well defined, due to the UVOT data, 
extending the photometric coverage to the near UV. 
If we assume a standard  Shakura \& Sunyaev (1973) disk
emission, the UVOT data permit fixing a lower limit to the mass of the black hole, 
which turns out to be around $10^8 M_\odot$, in agreement with the estimates by Zhou 
et al. (2003). For lower masses, in fact, the luminosity needed to fit these data 
becomes super--Eddington.

We model the contemporaneous optical to $\gamma$--ray data with a one--zone synchrotron and inverse 
Compton model, in accordance with the models generally used for blazars. 
Also for this source, the radio emission is assumed to come from larger--scale emission regions further 
away along the jet, while the rest of the SED is attributed to the region at the beginning 
of the jet plus the contribution of the accretion disk. 
The IR radiation produced by the 
assumed torus does not influnce the derived non--thermal SED, since the corresponding 
radiation energy density is much smaller than the one produced by the lines.

\begin{figure*}
\centering
\includegraphics[scale=0.4]{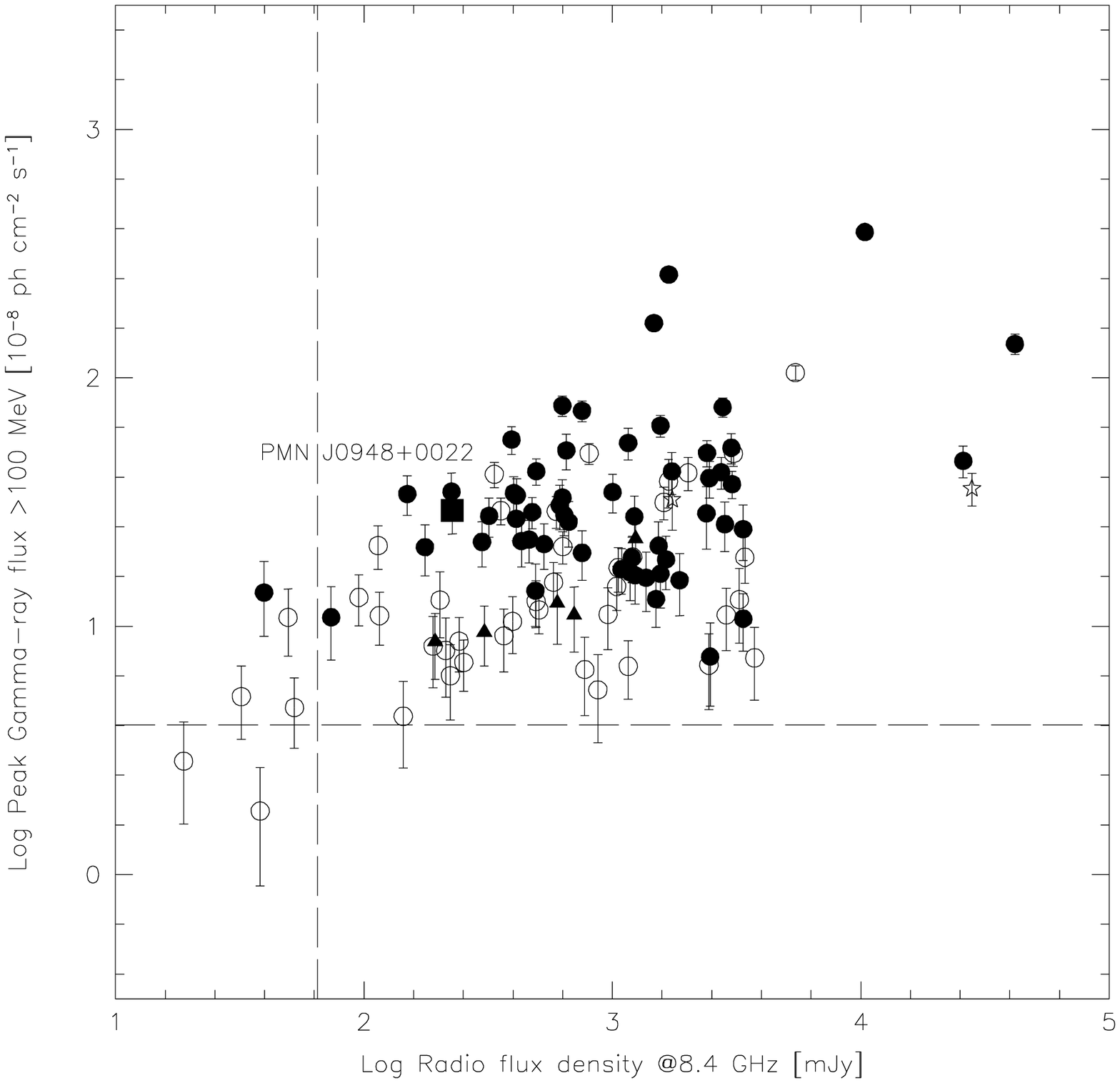}
\includegraphics[scale=0.4]{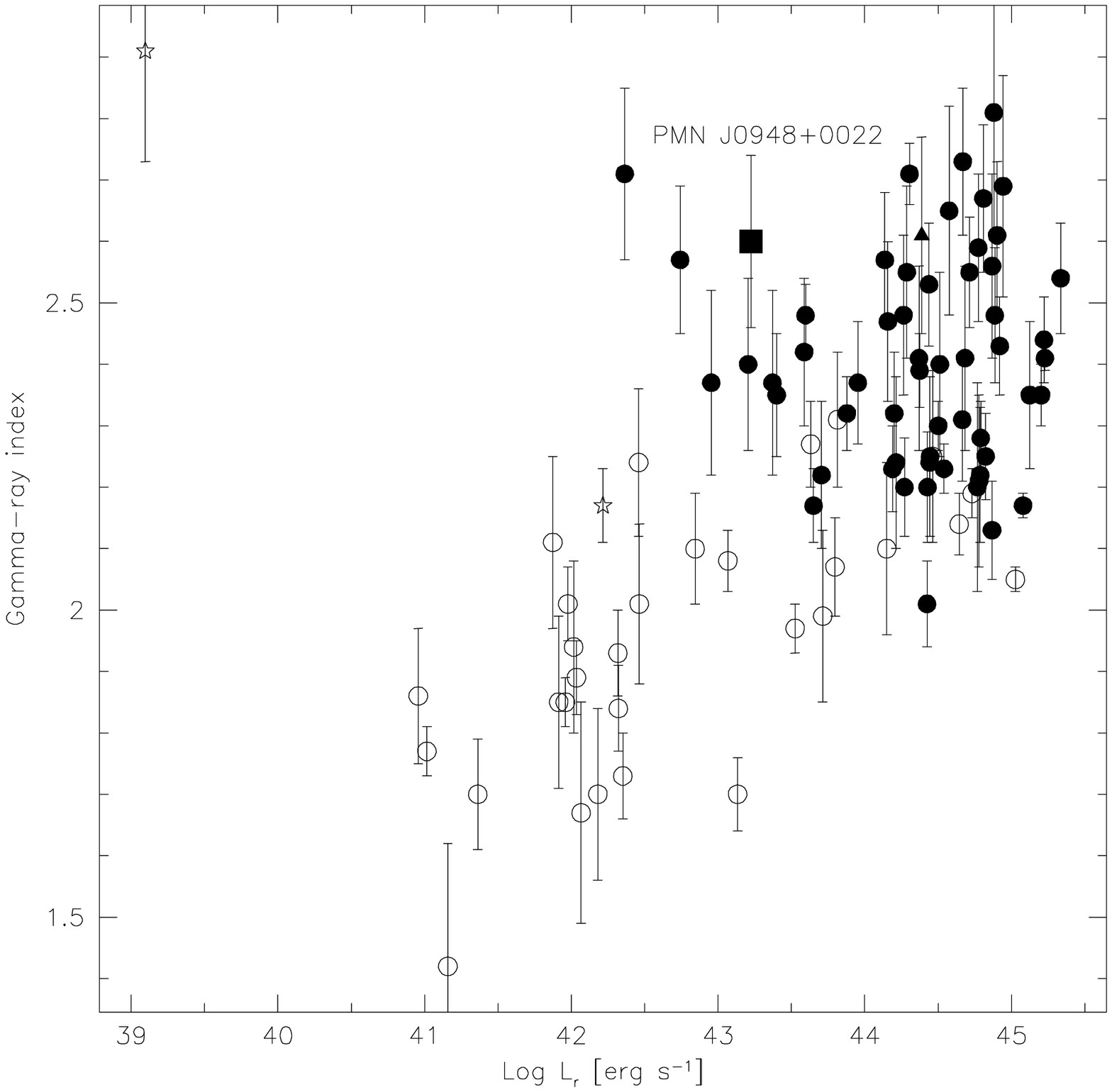}
\caption{Radio versus $\gamma$-ray properties of PMN~J0948+0022 (indicated with a filled square) compared with the other blazars detected by LAT (figure adapted from Fig. 14 in Abdo et al. 2009b). (\emph{left panel}) Peak $\gamma$-ray flux ($E>100$~MeV) vs. radio flux density at 8.4 GHz; the dashed lines show the CRATES flux density limit and the typical LAT detection threshold. (\emph{right panel}) $\gamma$-ray photon index vs. radio luminosity.}
\label{radiogamma}
\end{figure*}

The complete description of the general model used can be found in Ghisellini \& 
Tavecchio (2009); here we briefly summarize the main parameters: 

\begin{itemize}
\item the source is assumed to be located at a distance $R_{\rm diss}=6.75\times 10^{16}$ cm from 
a black hole of mass $M=1.5\times 10^8M_\odot$; 

\item the source is assumed to be a sphere, of radius $R=6.75\times 10^{15}$cm (i.e. a conical jet 
of semiaperture angle $\psi=0.1$~rad is assumed);

\item it moves with a bulk Lorentz factor $\Gamma=10$ at a viewing angle $\theta_{\rm v}=6^\circ$; 

\item the disk luminosity is 40\% of the Eddington value ($L_{\rm disk}=9\times 10^{45}$~erg~s$^{-1}$); 

\item above (and below) the disk we assume an X--ray emitting corona, producing a luminosity 
$L_{\rm cor}=0.3 L_{\rm disk}$ and with a power law spectrum (energy index $\alpha_X=1$), ending 
in an exponential cut at 150~keV;

\item 10\% of the disk emission is assumed to be absorbed and re--emitted by the broad-emission 
line region (BLR), which emits narrow permitted lines, in this specific case;

\item the BLR is placed at a distance $R_{\rm BLR}=10^{17}L_{\rm disk,45}^{1/2}=3\times 10^{17}$~cm, 
where $L_{\rm disk,45}$ is the disk luminosity in units of $10^{45}$~erg~s$^{-1}$; 

\item a dusty torus absorbs 10\% of $L_{\rm disk}$ re--emitting it in the far--IR. Its distance 
is assumed to be $R_{\rm IR}=2.5\times 10^{18} L_{\rm disk,45}^{1/2}$~cm, calculated by considering 
that the dust is at the temperature of 370 K (see Cleary et al. 2007), where is the maximum of 
efficiency in reprocessing the impinging photons into IR radiation.

\end{itemize}

For the above choice of parameters, the BLR is the main producer of the seed photons for the inverse Compton scattering in the 
emitting region. In Fig. \ref{sed}, this component is labeled EC. On the other hand, the 
Synchrotron Self--Compton (SSC) radiation is important in the X--ray band, where it 
dominates the radiative output (the two--peak shape of the SSC curve correspond to the 
first and second order scattering). In this scenario, i.e. with the X-ray data due to 
the SSC emission, the magnetic field $B$ is strongly constrained, since it controls the 
value of the SSC emission. In our case $B=3.2$ G. 
 
The particle energy distribution of the emitting electrons is calculated by the continuity 
equation, assuming the injected electrons are distributed in energy according to a smoothly 
joining broken power law of slopes $\gamma^{-1}$ and $\gamma^{-2.2}$ below and above  
$\gamma=800$, respectively. The maximum electron energy corresponds to a random Lorentz 
factor of $\gamma_{\rm max}=1600$ and the inverse-Compton peak corresponds to 
$\gamma_{\rm peak}=411$. The total power injected in the form of relativistic 
electrons is $L^\prime_{\rm e}=2.4\times 10^{43}$~erg~s$^{-1}$, as calculated 
in the comoving frame. 

Once accelerated, the bulk kinetic power carried by the electrons in the jet is 
$L_{\rm e}=5.0\times 10^{44}$~erg~s$^{-1}$, to be compared with the Poynting 
flux $L_{\rm B}=1.8\times 10^{44}$~erg~s$^{-1}$ and the power in radiation, 
$L_{\rm rad}=2.0\times 10^{45}$~erg~s$^{-1}$. 
As occurs in typical powerful blazars 
(see e.g. Celotti \& Ghisellini 2008, also for the exact definitions of these powers), 
the radiation observed carries more power than what is available in magnetic 
field and electron energy. 
Thus we require that the jet also carries protons: if 
we assume one proton per electron, we obtain $L_{\rm p}=4.8\times 10^{46}$~erg~s$^{-1}$. 
This value can be reduced, by assuming the presence of pairs.

The impact of the presence of narrow permitted lines in the BLR can be evaluated in 
two ways. 
First, there can be a geometrical explanation: if the BLR is torus--like 
and we are observing it face--on, then the Doppler broadening is reduced and the line 
width is smaller than usual. 
A similar geometry has been invoked by Decarli et al. (2008) 
to explain why NLS1 have not small masses. 

This BLR geometry, although different from what used in the
adopted model, does not affect any of the model results,
since the important parameter is the angle at which the jet sees the BLR. 
Second, it is still 
possible to consider a isotropic geometry of the BLR and take into account the effect 
of the radiation pressure, as proposed by Marconi et al. (2008). 
The BLR is then moved 
farther off the central singularity ($R_{\rm BLR}=5\times 10^{17}$~cm), but the impact 
in the model parameters mainly results in an increase of $\Gamma$ from 10 to 13. Other 
new values of parameters are: $\gamma_{\rm max}=1500$, $\gamma_{\rm peak}=430$ 
$L^\prime_{\rm e}=2.6\times 10^{43}$~erg~s$^{-1}$, $B=2.01$~Gauss. 
The power output 
of the jet is: $L_{\rm e}=1.2\times 10^{45}$~erg~s$^{-1}$, $L_{\rm B}=1.3\times 
10^{44}$~erg~s$^{-1}$, $L_{\rm rad}=3.2\times 10^{45}$~erg~s$^{-1}$, 
$L_{\rm p}=7.9\times 10^{46}$~erg~s$^{-1}$.

It is worth noting that the systematics in the LAT data do not affect significantly 
the results of the modeling. Indeed, in the case of a LAT flux equal to the 130\% 
of the observed value (the worst case), it is sufficient to increase a little the 
injected power $L^\prime_{\rm e}$ (from $2.4\times 10^{43}$ to $2.8\times 10^{43}$~erg~s$^{-1}$) 
and to decrease a little the magnetic field $B$ (from $3.2$ to $2.5$~Gauss). The derived 
quantities become $L_{\rm e}=6.3\times 10^{44}$~erg~s$^{-1}$, 
$L_{\rm B}=1.0\times 10^{44}$~erg~s$^{-1}$, $L_{\rm rad}=2.5\times 10^{45}$~erg~s$^{-1}$, 
$L_{\rm p}=5.0\times 10^{46}$~erg~s$^{-1}$. 
A change in the $\gamma$-ray photon 
index equal to $\pm 0.1$ results in very negligible changes in the SED.

The model parameters used to explain the SED of PMN~J0948+0022
are rather well constrained, given the basic assumptions of the model.
However, they may not be unique.
Another solution might be possible if we assume that the
dusty torus contributing to the IR seed photons is hotter and more compact 
than assumed here. It must be hotter because otherwise we need
too energetic electrons to fit the hard X--ray emission (by EC), 
and these very same electrons would overproduce (by synchrotron)
the observed emission in the optical.
A stratified and clumpy torus (e.g. Nenkova et al. 2008) could work.
Also in this "hot and clumpy torus" case 
we can obtain a reasonable representation of the data
(i.e. the same ratios between the radiation energy densities
 -- both by synchrotron and external photons -- and the magnetic energy density),
at distances one order of magnitude larger than assumed here.
As a consequence, the emitting region should be one order of magnitude
larger, and thus should vary on longer timescales.
The detection of a typical variability timescale can therefore 
discriminate between the two possible solutions.

\section{Discussion and Conclusions}

Our findings show clearly that PMN~J0948+0022, the first narrow-line quasar 
detected in $\gamma$--rays, hosts a relativistic jet with SED very similar to those 
of ``classical'' FSRQ.

The parameters relevant for the spectral shape of the SED derived from the modelling, 
using the approach by Ghisellini and Tavecchio (2009), are consistent with those of 
high--power blazars, though at the lower limit in power of the FSRQ region as derived, 
for example, in Celotti \& Ghisellini (2008) for a large sample of blazars 
(cf. their Fig.~2 and Fig.~6). The comparison of radio vs $\gamma$--ray properties 
shows that in both bands PMN~J0948+0022 has a relatively low power with respect to 
the other FSRQ detected by LAT to date, although no striking differences appear 
(Fig.~\ref{radiogamma}). 
Effelsberg and OVRO radio observations also show characteristics similar to those of FSRQ.

We believe that PMN~J0948+0022 could be one of the first examples of a FSRQ with relatively 
small mass but high accretion rate in terms of Eddington ratio. This would fit well with 
the scenario in which the transition between high peaked and low peaked blazars is not 
related to absolute power but to the Eddington ratio as indicated by the FRI-FRII 
separation which depends on mass. It remains to be seen whether the jets in other 
radio-loud NLS1 conform to this scenario. \emph{Fermi} is expected to answer this 
question by measuring the $\gamma$-ray luminosity and spectra of the most radio-loud ones.

\acknowledgments
The \emph{Fermi}/LAT Collaboration acknowledges generous ongoing support from a number 
of agencies and institutes that have supported both the development and the operation 
of the LAT as well as scientific data analysis.  These include the National Aeronautics 
and Space Administration and the Department of Energy in the United States, the 
Commissariat \`a l'Energie Atomique and the Centre National de la Recherche Scientifique 
/ Institut National de Physique Nucl\'eaire et de Physique des Particules in France, 
the Agenzia Spaziale Italiana and the Istituto Nazionale di Fisica Nucleare in Italy, 
the Ministry of Education, Culture, Sports, Science and Technology (MEXT), High Energy 
Accelerator Research Organization (KEK) and Japan Aerospace Exploration Agency (JAXA) 
in Japan, and the K.~A.~Wallenberg Foundation, the Swedish Research Council and the 
Swedish National Space Board in Sweden.

Additional support for science analysis during the operations phase from the following 
agencies is also gratefully acknowledged: the Istituto Nazionale di Astrofisica in 
Italy and the K.~A.~Wallenberg Foundation in Sweden for providing a grant in support 
of a Royal Swedish Academy of Sciences Research fellowship for JC.

This research is partly based on observations with the 100-m telescope of the MPIfR 
(Max-Planck-Institut f\"ur Radioastronomie) at Effelsberg. The monitoring program at 
the OVRO is supported by NASA award NNX08AW31G and NSF award AST-0808050.

This research has made use of the NASA/IPAC Extragalactic Database (NED) which is 
operated by the Jet Propulsion Laboratory, California Institute of Technology, 
under contract with the National Aeronautics and Space Administration. This 
research has made use of data obtained from the High Energy Astrophysics 
Science Archive Research Center (HEASARC), provided by NASA's Goddard Space Flight Center.

\end{document}